**Cognitive information encryption in the expert management of strategic uncertainty**

Seth Frey[a,b,1], Paul L. Williams[b], Dominic K. Albino[c,2]

[a] Disney Research Zürich, Stampfenbachstrasse 48, 8006, Zürich, Switzerland
[b] Cognitive Science Program, Indiana University, 1900 E 10th St., Bloomington, IN, 47406, USA
[c] Department of Economics, University of Connecticut, 365 Fairfield Way, Storrs, CT 06269, USA

[1] Correspondent: seth.frey@dartmouth.edu

[2] Former professional No-Limit Hold'em cash game player.

**Abstract**

Strategic agents in incomplete-information environments have a conflicted relationship with uncertainty: it can keep them unpredictable to their opponents, but it must also be overcome to predict the actions of those opponents. We use a multivariate generalization of information theory to characterize the information processing behavior of strategic reasoning experts. We compare expert and novice poker players — "sharks" and "fish" — over 1.75 million hands of online two-player No-Limit Texas Hold'em (NLHE). Comparing the effects of privately known and publicly signaled information on wagering behavior, we find that the behavior of sharks coheres with information that emerges only from the interaction of public and private sources — "synergistic" information that does not exist in either source alone. This implies that the effect of public information on shark behavior is better encrypted: it cannot be reconstructed without access to the hidden state of private cards. Integrative information processing affects not only one's own strategic behavior, but the ability of others to predict it. By characterizing the informational structure of complex strategic interactions, we offer a detailed account of how experts extract, process, and conceal valuable information in high-uncertainty, high-stakes competitive environments.

When valuable information is common knowledge, utility-maximizing strategic agents face a difficult problem. Failing to use the information means forgoing an opportunity to exploit opponents, but using it can make one more predictable to opponents. How do expert strategic reasoners manage the competing demands of maximizing utility and minimizing predictability? In games of incomplete information, is the choice between these goals inherently a tradeoff, or is it possible to have one without losing the other?

Game experts are an excellent source of insight into how humans process strategic information in competitive settings. Interest in strategic experts, and chess experts in particular, goes back to the very roots of both cognitive science and artificial intelligence (Chase & Simon, 1973). More recently, chess experts have attracted the attention of behavioral economists (Bühren & Frank, 2012; Gerdes, Gränsmark, & Rosholm, 2011; Levitt, List, & Sadoff, 2011), who have found, for example, that grandmasters exhibit impressively many levels of recursive strategic reasoning when playing simple economic games against each other, and that they adapt to novices by exhibiting fewer levels of recursive reasoning (Palacios-Huerta & Volij, 2009).

Though the complete-information game of chess has received the most attention, theorists have been making progress on incomplete-information games since the 1950's, with a special focus on subgames of poker (Billings, Davidson, Schaeffer, & Szafron, 2002; Kuhn, 1950). In a recent major advance, Bowling et al. present a type of solution to two-player limit Hold'em (Bowling, Burch, Johanson, & Tammelin, 2015; Sandholm, 2015). Subgames of poker have also proven useful for the experimental study of human strategic behavior (Laakasuo, Palomäki, & Salmela, 2014; Meyer, Meduna, Brosowski, & Hayer, 2013; Rapoport, Erev, Abraham, & Olson, 1997; Seale & Phelan, 2009). As econometric methods are brought to bear in field settings, academics are starting to build a consensus that poker is a game of skill, at both low and high stakes (Croson, Fishman, & Pope, 2008; Levitt & Miles, 2014; "Poker as a Skill Game: Rational vs Irrational Behaviors," 2014; Potter van Loon, van den Assem, & van Dolder, 2014). But our understanding of information processing strategies in complicated, uncertain environments remains limited. Also limited is our understanding of the role of multivariate interactions in group dynamics and strategic reasoning processes. Though researchers have begun to model interactions in simplified, simulated team settings (Oliver & Fienen, 2009), our work offers a general approach for estimating multivariate interactions in large behavioral datasets.

Poker in its pure form is a zero-sum game. But in professional settings, and in our dataset, the game's hosts take small fees, making it negative-sum. No-limit Texas Hold'em (NLHE) is a particularly relevant variant of poker for those interested in strategic information processing under incomplete information. One reason for this is that it is designed to make bluffing a central aspect of play: players have relatively few observable behaviors available to them, each of which is capable of signaling the strength of a player's private state, whether honestly or dishonestly.

In each hand of Hold'em, players receive their two private "hole cards" before proceeding through four voluntary rounds of betting on the value of those cards. Each hand opens with two players posting mandatory small wagers, commonly called "blinds." The blind establishes the table's betting level. Because it defines the minimum bet or raise, it is commonly interpreted as a proxy of the table's stakes, and also its competitiveness, with higher-stakes tables attracting more skilled players. In each betting round, players act in sequence around the table, choosing either to "fold" and quit the hand, or to continue. Betting rounds are punctuated with the incremental addition of community cards (ultimately five), that are common knowledge and qualify as being in the hands of all players. Players who quit a hand lose any money they contributed and wait for the next hand to begin. Players who continue must wager enough to either match or exceed the current highest bet. If all players but one fold, the remaining player wins all money contributed during that hand, the "pot," without showing his or her cards. If there are still contenders after the final betting round there is a "showdown." Hands are valued based on how their cards relate to each other, and the player with the highest-valued hand wins the pot. Only during showdown do players reveal their cards, and often only the player who ended up winning the hand. The game offers many mechanisms by which players can strategically misinform each other about the value of their cards. Players with strong hands may signal weak hands with small bets to keep the pot growing, and players with weak hands may signal strong hands with large bets to intimidate their opponents into folding before showdown.

The online setting that we focus on introduces a few differences from traditional forms of the game. Eliminating physical co-presence as a feature removes many sources of information about players (their "tells"), but it also creates new ones. Most online experts build behavioral dossiers on their opponents, using real-time analysis software to collect, and even buy, records of other players' "hand histories." With this software, information about opponents arrives quickly and in a format ready for automatic analysis.

It is common to distinguish between expert poker players, "sharks," and the remaining "fish." We make the distinction in terms of players' mean profits per hand, defining as sharks players whose mean earnings are positive, and fish as the least profitable 50% of unprofitable players. In our dataset, profitable players constitute just 10–15% of the total population. For narrative convenience, we will frame our presentation around a focal player, the female "hero," playing in competition with a male opponent, the "villain."[1]

Our empirical approach is based on estimating, from all two-player hands, the information about the hero's wagering behavior — random variable *W1* — that is revealed by the private strength of her hand (*P1*) and by the villain's wagers (*W2*). We apply an information theoretic measure to decompose the types of influences on *W1* from *P1* and *W2,* interpreting *P1* as a private information channel and *W2* as a public information channel.

Partial information decomposition is a method that was developed to generalize the core concepts of information theory to the analysis of multivariate interactions (Beer & Williams, 2014; Williams & Beer, 2010; 2011). It represents a recent advance toward making information theory practical for the analysis of high-dimensional systems. More established measures like interaction information, transfer entropy, and multi-information share the feature that, above two variables, they characterize some relationships in terms of negative quantities of information (Cover & Thomas, 2006; McGill, 1954; Schreiber, 2000). Not only does this make them difficult to interpret, it is symptomatic of a deeper confound between types of higher-order interactions that partial information decomposition, the more recent method, is capable of distinguishing (Williams & Beer, 2010). Rather than framing its analysis in terms of a generalization of mutual information, partial information decomposition operates on the "total information," $I(X;Y)$, of variables *Y* "about" a focal output variable *X*. This "aboutness" comes from the causal structure that must hold for partial information decomposition to be effective: the variables *Y* must be the inputs upon which *X* operates. Total information is the statistical coherence that an output variable shares with every subset of input variables *Y*. This is the sense in which it is the total of the information about $X$ contained anywhere else in the system. It is relatively simple to calculate as a sum of mutual information terms; for three variables it is the sum of the interaction information (McGill, 1954) and two pairwise mutual information terms,

$$I(X; Y_1, Y_2) = I(X; Y_1; Y_2) + I(X; Y_1) + I(X; Y_2)$$

______________________

[1] For political convenience, genders were assigned by coin flip.

Partial information decomposition provides an alternative specification of this total information in terms of non-negative components with clear interpretations,

$$I(X; Y_1, Y_2) = Unq(X; Y_1) + Unq(X; Y_2) + Rdn(X; Y_1, Y_2) + Syn(X; Y_1, Y_2)$$

The four terms of this specification correspond, in turn, to the information about $X$ from each of $Y_1$ and $Y_2$ *uniquely* (*Unq(X;Y_1)* and *Unq(X;Y_2)*), the information coming *redundantly* from both (*Rdn(X;Y_1,Y_2)*), and the *synergy*: the information about $X$ from both $Y_1$ and $Y_2$ that could not have come from either alone (*Syn(X;Y_1,Y_2)*). The synergy can be seen as information that emerges from the interaction of the input variables. See the Methods section for definitions of these terms.

As a technique for decomposing multi-dimensional interactions, partial information decomposition is of growing interest in the study of complex systems (Bertschinger, Rauh, Olbrich, Jost, & Ay, 2014; Griffith & Koch, 2014; Griffith, Chong, James, Ellison, & Crutchfield, 2014; Harder, Salge, & Polani, 2013; James, Ellison, & Crutchfield, 2011), particularly in theoretical neuroscience (Flecker, Alford, Beggs, Williams, & Beer, 2011; Lizier, Flecker, & Williams, 2013; Timme, Alford, Flecker, & Beggs, 2013), though, to-date, it has only been used to analyze the results of simulated neural and computational systems. Applied to "real" human behavioral data, in our case the signaled and actual hand strengths determining a poker player's wagering behavior, partial information decomposition can elucidate how players manage the advantages and disadvantages of using public information sources. For example, synergistic information about *W1* that came from the interaction between *P1* and *W2*, and not from either alone, must have come as a result of the hero's strategic calculations, allowing us one measure of the extent or efficacy of her strategic reasoning processes.

Our analysis starts with the distribution over discretized values of *W1*, *P1*, and *W2*. We defined *W1* and *W2* by discretizing the range of observed wagers into three states, *No wager*, *Small wager*, and *Large wager*. We defined *P1* by discretizing all possible hand strengths into the states *Weak hand*, *Strong hand*, and *Not observed*. The details of our discretization procedure are explained in the Methods section. We then used these values to populate multiple *3×3×3* distributions, calculating separate distributions based on whether the hero was a fish or a shark, as well as for different betting levels. The conventional wisdom is that tables at higher betting levels attract more skillful players. Though our conclusions do not depend on the validity of this impression, we recalculated all results within betting level to control for the possibility that playing styles change by blind. To measure the significance of observed differences across the different joint distributions, we bootstrapped 95% confidence intervals for each informational measure. We used the resulting values of statistical coherence between players' signals to ground our inferences about expert information use.

Though causal statements are difficult to justify in observational datasets, the rules and incentives of poker add constraints and permit assumptions that we can use to structure our inferences. We make a number of assumptions: online poker players are motivated to maximize earnings; NLHE is a game of skill; profitable players are particularly skillful, more so than unprofitable players; higher stakes imply a higher level of competitiveness in play and higher skill among the best players at a table (if not all of them); and sharks are more likely than fish to have informative priors about their opponents' skill levels — at least because they are more likely to use analysis software. We detail and defend our assumptions in the Methods.

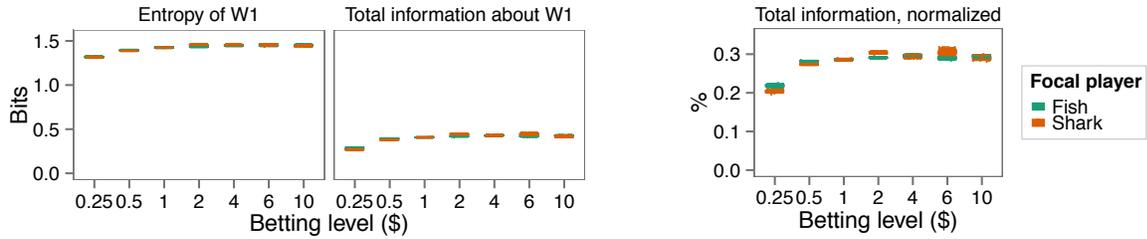

**Figure 1. Entropy of *W1* and total information about it, in bits and normalized.**

Across both the raw predictability of a wager (Information entropy, *H(W1)*) and its predictability as a function of private and public information sources (Total information, *I(W1;P1,W2)*), there is no difference between poker experts and novices, between "sharks" and "fish." To provide another view into the relative predictability of the hero's wagers, the third panel plots the quotient of the data in the second panel over that in the first — total information as a fraction of entropy (*I(W1;P1,W2))/H(W1)*). Whatever the difference between shark and fish wagering behavior, it does not seem to be due to differences in the quantity of bits extracted from hole cards or the wagers of others.

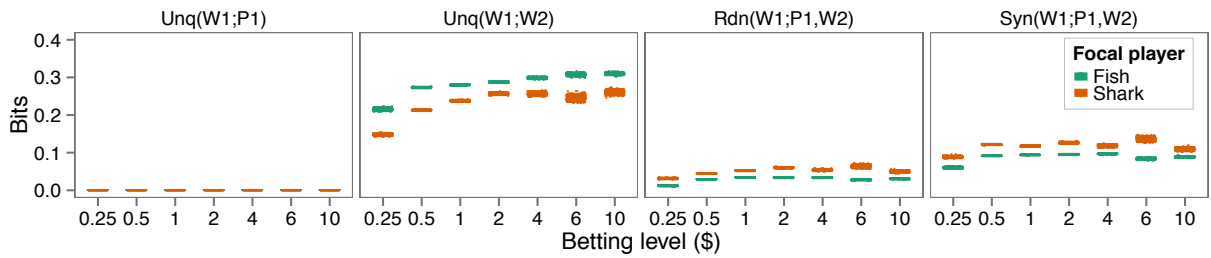

**Figure 2. Information decomposition of *I(W1;P1,W2)*. Sharks show greater synergy than fish.** As compared to those of novice "fish," the wagers (*W1*) of expert "shark" players exhibit significantly higher levels of synergy, and significantly lower amounts of unique information from the "villain" opponent's wagers (*W2*). Unique information from the focal "hero" player's cards (*P1*) is zero in part because coherence between *W1* and *W2* is particularly high in two-player games.

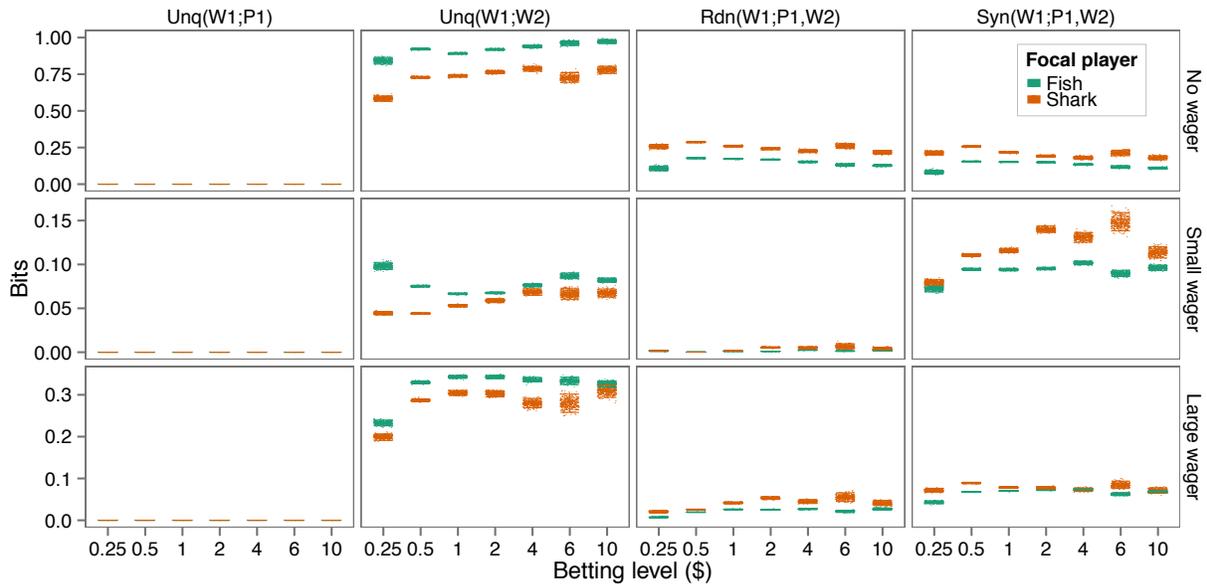

**Figure 3. Specific information decomposition of the total information, *I(W1;P1,W2)*.** Sharks exhibit significantly more synergy for all states of *W1*, particularly when they made small wagers or no wager. The unique information from the villain's wagers is a particularly strong predictor that the hero will place no wager. The specific information decomposition is defined in the Supplementary Information (SI).

**Results**

Fig. 1 plots three measures of the net predictability of the hero's (focal player's) wagers, including the total information about them (second panel). The four panels of Fig. 2 decompose the total information into its four parts. Fig. 3, in turn, further decomposes Fig. 2, with a separate information decomposition for each state of *W1*. Unless otherwise labeled, all panels plot the number of bits of an information theoretic measure, and are partitioned along the horizontal axis by betting level (blind). Confidence intervals in all graphs were calculated by resampling the data 500 times. Across measures, a wide interval is as likely due to low data availability as to intrinsic variance. For example, since sharks are less common than fish, and since high-stakes games are less common than low-stakes games, the intervals for games with sharks or with high stakes tend to be wider. Table S1 reports descriptive statistics, including the quantities of data and the binning cutoffs per betting level.

### ***Two conflicting senses of predictability with an increase in table competitiveness***

How do poker players align themselves to available information without becoming predictable themselves? To answer this question we first identify two senses of "predictability." In all three panels of Fig. 1 there is less information about the relevant variable at 25¢ tables than at higher stakes. And yet, the implication of that result in the first panel is in some sense opposed to its implication in the other two: the lower entropy signaled in the first panel implies that wagers at low-stakes tables are *less* predictable, while the other two imply that they are *more* predictable. One interpretation of this discrepancy is that behavior at higher stakes is more complex, in the narrow sense that it takes more information to describe, and that it is simultaneously more deterministic, in the narrow sense that game state variables predict a greater share of behavior. To be precise, it takes more bits of information to describe *W1* at higher stakes (Fig. 1, panel 1) but, given $P1$ and $W2$, more of that information can be predicted (panels 2 and 3).

### Sharks do not extract more bits of information than fish

Sharks and fish show no difference in the total information about $W1$ (Fig. 1, panel 2). As elaborated in the next subsections, fish's wagers have more statistical coherence with uniquely public information and sharks' wagers cohere more with the redundancy and the synergy between both information sources (Fig. 2). Sharks seem to distinguish themselves not in the number of bits they are able to extract from the game's information channels, but in how they process and manage those bits.

### Fish's wagers have more statistical coherence with public information uniquely

Across all betting levels, sharks and fish exhibit significant differences in their values of $Unq(W1;W2)$ (Fig. 2, panel 2). This means that more information in the wagers of fish is derived exclusively from information in the villain's wagers, or from public information that could have influenced both players. The implications of this result depend on the assumptions that one is comfortable making. For example, if one allows that informational differences between sharks and fish point to areas of strategic advantage for the former, we could interpret this result as indicating that fish cohere too closely with public signals uniquely.

### The effects of opponents' wagers on a shark's behavior are encrypted

As compared to fish, more information about the wagers of sharks is contained in the synergy between $P1$ and $W2$, $Syn(W1;P1,W2)$ (Fig. 2, panel 4). Synergistic information about an output variable is the information that comes from the two input variables' interactions, and not from either alone. As the only person with access to both sources of information, the interaction could only have occurred in the mind of the hero, and we can conclude that sharks perform more integrative processing of the input variables in calculating their wagers. And because synergy reflects information that can only be gained with access to both input variables, the full effects of one cannot be reconstructed without access to the other. Since one of the inputs to the synergy, *P1*, is known only to the hero, information in this component is encrypted; even though *W2* is common knowledge, any synergistic contribution it makes to *W1* cannot be inferred as long as *P1* remains unknown. It should come as no surprise that experts are more integrative information processors; it is entirely likely that more processing leads to more profitable behavior in complex strategic settings like NLHE. The surprise is the ability of synergistic information processing, beyond its direct effect on one's own behavior, to increase opponents' strategic uncertainty about what that behavior will be.

### ***Other results: Skill across stakes***

Across both Figs. 1 and 2, the amount and distribution of information at 25¢ tables differs greatly from that at every other betting level. Additionally, those other betting levels do not seem to differ from each other. According to the experiences of author DKA, a former professional online Hold'em player, the differences in skill between profitable players in different blinds are very small at and above 50¢. While profitable players at 25¢ tables may still be considered novices, a player who is profitable at even a 50¢ table can be considered to be an overall excellent poker player. The validity of our analysis does not rely on the validity of this impression, but our results are consistent with its implications.

### ***Other results: Specific partial information decomposition***

The results illustrated by Figs. 1 and 2 have focused on the average total information, over all states of the hero's wager. However, the partial information decomposition can also be applied at a finer-grained level, over each of the three states of *W1: No wager*, *Small wager*, and *Large* *wager (*accounting for about 15%, 50%, and 35% of wagers, respectively; Table S1*)*. The three rows of Fig. 3 break Fig. 2 into three separate information decompositions. Fig. 3 shows how unique, redundant, and synergistic effects of *P1* and *W2* differ in their power to predict whether the hero made a large, small, or zero wager. The increased synergy that sharks exhibit before wagering zero (top row) may represent skill at reducing losses, while the synergy determining their large wagers (bottom row) may correspond with their efforts to maximize gains. Sharks seem to distinguish themselves with particularly high levels of synergy in the cases of zero and small wagers (top and middle rows).

### ***Other results***

The first panel of Fig. 2 suggests that there is no unique contribution of *P1* to *W1* under any conditions. Why might it be that *Unq(W1;P1)* is always zero? By elaborating on this finding, we clarify a few points about interpreting the results of a partial information decomposition. First, the absence of a unique contribution of *P1* does not imply a general lack of information from that source. The definitions of unique and redundant information, in the Methods section, show that there will be no unique contribution from one input variable if the other always gives at least as much information about each state of the output variable. This is essentially guaranteed to be the case in two-player NLHE. First, the ubiquity of bluffing in NLHE ensures relatively low statistical coherence between *P1* and *W1*. Second, the two-player setting guarantees relatively high coherence between *W2* and *W1*. For example, only among two players does one player's fold necessarily imply a win for someone else.

In the third panel of Fig. 2, *Rdn(W1;P1,W2)* is significantly higher among sharks. The rise of redundancy *may* reflect more coherence of sharks' wagers with *W2*, but because *Unq(W1;P1)=0* it *must* at least reflect an increase in their coherence with *P1*. Speculatively, the higher coherence of sharks with their cards may reflect their greater skill at extracting information from their private hole cards.

### *Robustness checks*

There were a number of ways in which we could have conducted this analysis. In our main analysis, we have restricted our attention to the player in second position and used all hands and all betting rounds in a hand. The risk of this inclusive design decision is that we allowed statistical dependencies between betting rounds and hands to contaminate our inferences. By restricting our analysis further, to behavior in only the first betting round, we could have eliminated unobserved influences of betting rounds on each other, with the downside of constraining the scope of our analysis to just this "preflop" play. Alternatively, by attending only to hands that ended with the revelation of hole cards during showdown, we could have reduced the statistical dependence of hands upon each other (because hands that end in showdown are infrequent), and also eliminated all unobserved values of P1, all at the cost of increased sampling bias and a much-reduced quantity of data. Last, by using data from players in both positions, we could have expanded the generality of our conclusions beyond second-position play, with the downside of obscuring our ability to infer unidirectional causal influence of *W2* and *P1* on *W1*.

Observing that the flaws of each of these analyses are compensated by an advantage of one of the others, we performed all four. Individually, each of these analyses either contributes to violations of statistical independence or restricts itself to a narrow and biased subset of the data. But taken together, they corroborate each other and collectively support the robustness of our findings to various violations of our core assumptions. The main text presents the results of the first analysis described above, and the results of the three variations are described in full detail in the Supplementary Information (SI).

Though there are many small differences in the results of each of these analysis, they are all in agreement that sharks exhibit greater levels of synergistic processing of *W2* and *P1* as inputs into their wagers. That all of these analyses are in strong qualitative agreement supports our major finding, that sharks draw a greater share of information from the interaction between public and private game information, thereby effecting strategic information encryption

**Discussion**

If a player's only goal were to be unpredictable, it would be enough to make her behavior a function of private information alone, or even to disregard all inputs, forgo all profits, and behave completely randomly. But these alternatives incur the opportunity cost of discarding valuable, exploitable information in an opponent's wagers. So how can sharks deterministically incorporate public information into their behavior — apply effective algorithms — without becoming predictable to other players?

Expert poker players distinguish themselves by extracting more information from how *W2* and *P1* interact. Since the latter of those sources is typically secret, the effects of public information on expert behavior will be encrypted; the influence of synergistic information on a wager cannot be reverse-engineered without access to both channels. Sharks can use public information without becoming predictable because the information in their cards acts like the "private key" in a public key cryptographic system. This result may constitute behavioral evidence for the use of "safe" strategies, which were recently introduced by computational game theorists. Nash equilibrium strategies, because they assume equilibrium behavior from others, can fail to exploit non-equilibrium "irrational" behavior. Safe strategies are non-Nash strategies that exploit irrational behavior in a way that is not vulnerable to exploitation (Ganzfried & Sandholm, 2012).

Sharks may not explicitly think of themselves as encrypting information; they may merely be performing more integrative computations over available information. If this is true, then the encryption of strategic information that we reveal is a mere side-effect, possibly unintended, of sharks' more thorough information processing behavior. Consequently, integrative information processing in Hold'em, beyond it's more obvious function of producing context-sensitive behavior, may serve the strategic function of obfuscating the mechanisms driving that behavior. We do not require the more daring proposal that poker experts are intentionally encrypting public information with the private information in their hole cards, but we do offer one piece of evidence in favor of that claim. If shark information processing were serving the most obvious alternative purpose, of extracting more bits of information from the available sources, sharks would be exhibiting higher levels of total information. According to the second panel of Fig. 1, they do not. Where they differ from fish is rather in how they allocate the same number of bits across unique, redundant, and synergistic components. Specifically, bits that fish extract from $W2$ uniquely are bits that sharks extract from the redundancy and the synergy between $W2$ and $P1$. That a relatively higher share of total information is coming from synergy is consistent with the hypothesis that strategic information encryption is a goal of expert information processing.

Information encryption is not a foolproof method for disguising behavioral patterns. The hero reveals her private state when she wins hands at showdown. With enough of these disclosures of her private information, it should be possible for the villain to reconstruct the algorithm she uses to compute her wagers from inputs $P1$ and $W2$. From a learning perspective, showdown is important because it forces players to expose data about how they integrate public and private channels. Thus, sharks that won the pot at showdown effectively sold information about their algorithm. Future work may be able to use such transactions to quantify the monetary value that players place on their own reasoning processes.

**Conclusion**

Professional poker players are experts at extracting signal from noise across many channels, and at integrating information from those channels both to exploit their opponents and protect themselves. Our analysis decomposes the empirically observed interactions within the three-dimensional system of a hero's public wagers, her private cards, and the public wagers of a single opponent. We suggest that poker sharks are more predictable than fish given all information (i.e. their behavior follows more predictably from game inputs), but that they are less predictable given only information observable to other players. Sharks accomplish this feat by carefully managing how their wagering behavior is informed by public and private channels. Their peculiar information management behavior allows them to maximize their coherence with the information they use without "tipping their hands."

In high-stakes competitive environments, success is not just about playing your cards right, but also playing your opponents right. Experts in such domains must extract information from others without revealing too much themselves. But greater influence of an opponent's behavior on one's own means more statistical coherence with it. How do strategic game experts stay unpredictable without decoupling themselves from the valuable information revealed by others? We show that poker experts pull their informational advantage not from their own cards, and not from their opponent's signals, but specifically from how those two information sources interact. Experts create new information by integrating public and private sources. Consequently, their behavior is encrypted: an opponent cannot reverse engineer the effects that their signals had on an expert without access to the "private key" information in that expert's cards. Expert strategic reasoning involves information encryption.

**Materials and Methods**

### *Measures*

Partial information decomposition generalizes the core concepts of information theory to the analysis of multivariate interactions. We describe it here starting with classic pairwise mutual information,

$$I(X;Y) = H(X) + H(Y) - H(X,Y)$$

where, for multivariate distribution **X,** the Shannon information entropy,

$$H(\mathbf{X}) = -\sum_{\mathbf{x} \in \mathbf{X}} p(\mathbf{x}) \log_2 p(\mathbf{x})$$

is a measure of the uncertainty of a system. Its unit is the bit. Mutual information functions as a generalized correlation, capturing arbitrary patterns of statistical coherence. With infinite data, entropy reflects any structure in a random variable, and mutual information detects any statistical coherence between variables, no matter the function that relates them. An alternate specification of mutual information is as the average *specific information*,

$$I(X;Y) = \sum_{x \in X} p(x) I(X = x; Y)$$

where

$$I(X = x; Y) = \sum_{y \in Y} p(y|x) \log(\frac{p(y|x)}{p(y)})$$

Specific information quantifies variable *Y*'s information about a specific state, *x*, of random variable *X.* The specific information will prove useful in our other definitions.

Although partial information decomposition is general enough to characterize multivariate interactions in arbitrarily high-dimensional distributions, our analysis is based on the simplest interesting case of three variables. The *total information* of two predictor variables about a focal output variable $X$ is

$$I(X; Y_1, Y_2) = I(X; Y_1; Y_2) + I(X; Y_1) + I(X; Y_2)$$

Its decomposition into four non-negative terms,

$$I(X; Y_1, Y_2) = Rdn(X; Y_1, Y_2) + Syn(X; Y_1, Y_2) + Unq(X; Y_1) + Unq(X; Y_2)$$

starts with *redundancy*, defined as the *minimum information*, $I_{min}$:

$$Rdn(X; \mathbf{Y}) = I_{min}(X; \mathbf{Y}) = \sum_{x \in X} p(x) \min_{Y \in \mathbf{Y}} I(X = x; Y)$$

$I_{min}$ is the expected value of the minimum specific information available from all predictors. In other words, it is the quantity of bits offered redundantly by all channels.

In three variables, the unique information between the output variable and an input is their mutual information minus any redundant contribution from the other input,

$$Unq(X, Y_1) = I(X; Y_1) - Rdn(X; Y_1, Y_2)$$
$$Unq(X, Y_2) = I(X; Y_2) - Rdn(X; Y_1, Y_2),$$

and the synergy is the amount of information remaining when the redundancy and the two unique components have been subtracted from the total information.

Unlike other information theoretic approaches to multivariate systems, the synergistic, redundant, and unique components of the partial information decomposition are strictly non-negative and simple to interpret. With this theoretical underpinning, it is possible to infer nonlinear coherences between arbitrarily many variables and to fully characterize how those variables complement and reinforce each other, down to the limits of discretization.

### *Data*

We used a popular dataset of hands collected by the academic community over about three weeks in 2009.[2] We restricted our analysis to the subset of this data from http://www.pokerstars.com. We parsed the data using the libraries of freepokerdb.[3] It failed to parse about 5% of hands.

Our base dataset included NLHE play from tables of all sizes (2–9 players; 9,274,066 hands). We used that data to calculate mean earnings per hand for all players and to calculate the binning boundaries for *W1*, *P1*, and *W2*. We restricted subsequent analyses to two-player tables (1,750,233 hands) over betting levels ("big blinds") of $0.25, $0.50, $1, $2, $4, $6, and $10. For each betting round of each hand, the data includes all publicly observable information: obfuscated but unique player IDs, community cards, blinds, wager sizes, pots, and the hole cards of any hands that were revealed at showdown. See Table S1 also.

### *Data structure*

We coded sharks as players whose mean profit per hand was positive. This accounted for only about 10–15% of players (See Table S1 for fraction of sharks per betting level). There may be noise in our operationalization of skill: either "true" experts who lost money in the data interval or "true" novices who were profitable. To manage for this concern we made the conservative decision to code as fish only the worst-performing half of the unprofitable 85–90% of players. Recognizing that unprofitable players at high-stakes tables may be more skillful than

---

profitable players at low-stakes tables, we gave players who played across betting levels a separate ranking at each level.

To support our causal claims, the focal hero player was defined in our main analysis as the second, responding, player at the table.[4] Since play order in NLHE rotates every hand, two players playing a series of hands would play a roughly equal number of them in the second position.

The empirical distributions at the foundation of our analysis consisted of three dimensions:

*W1*: We used the hero's **wagers** as a signal of the strength of her hand, with *W1* defined as a random variable over the states *{No wager, Small wager, Large wager}*. *W1* is the focal variable for our analysis; it is the hero's output after integrating inputs *P1* and *W2*.

— **P1**: The **private** or intrinsic strength of the hero's hand, given both hole cards and community cards available at a given betting round. In our main analysis, the possible states of *P1* were *{Not observed, Weak hand, Strong hand}*.

— **W2**: The wagers of the villain, with the same states as *W1*. *W2* also functions as a proxy for exogenous events that could have influenced both players.

We discretized signaled-hand-strength variables *W1* and *W2* using equal-frequency bins, in which bin boundaries are adjusted to try to ensure an equal number of observations in each bin. An equal-frequency binning allows for the most conservative and uninformative discretization of a variable, since it makes fewer assumptions about the underlying distribution than any other possible binning. Because they both signal weak cards, we coded as *No wager* both folds and "checks" — special bets of zero. Of the remaining wagers, the top half of non-zero wagers within a given blind were coded as *Large*, and the bottom half as *Small*. This binning incidentally followed implicit norms of the game: the *Small*/*Large* cutoff within each stake level was always equal to the stakes (and wagers equal to the cutoff were always binned as *Small*): at $2.00 tables, every wager strictly greater than $2.00 was coded as *Large*.

---

[4] To accommodate potential downsides of this modeling decision, the third robustness check in the SI reproduces our results in the case that the hero is not restricted to a particular table position.

To discretize the intrinsic strength of the hero's hand into variable *P1*, we first constructed an ordinal scoring system with the property that a particular hand dominates all hands with a smaller score. The scoring of *P1* was not based only on the hero's hole cards, but on how those cards combined with the community cards.[5] We then discretized the ordinal ratings into a binary *Weak*/*Strong* rating by binning within each betting level, over all betting rounds. Out of all observed hands at a betting level, the top half were coded as strong and the bottom half as weak. The cutoff usually designated everything above a low pair as *Strong* (Table S1). Because our dataset only included publicly observed behavior, and because most hands end before showdown, most values of *P1* were *Not observed*.[6]

### *Causal inference*

The rules and incentives of poker help satisfy assumptions that add causal structure to our inferences. We interpret the decomposition of *I(W1;P1,W2)* in terms of the effects of inputs *P1* and *W2* on output behavior *W1*. *P1* is self-evidently a unidirectional influence on *W1*. The relationship between *W1* and *W2* is more complicated, and there is room for mutually causal influences at three scales: within a betting round, across betting rounds within a hand, and across hands. Still, we argue that any synergy or redundancy *about* one of these three variables *from* the other two must be about *W1* from *P1* and *W2*, since *W1* is the only of the variables with causal inputs from both of the others:

1.  There cannot be synergy between *W1* and *W2* affecting *P1* because there is no causal link from *W2* or *W1* to *P1*. Cards are dealt before players begin wagering. *P1* is also independent of events or information in previous hands.

2.  There cannot be synergy between *W1* and *P1* about *W2*. This is partly because of the nature of *P1*. The only possible links from *P1* to *W2* are via cheating, *W1*, or *P2* (the villain's hole cards), none of which can foster synergy between *W1* and *P1* about *W2*.

---

[5] Our scheme included no sense of the speculative value of a hand, the increased likelihood for certain weak hands to become strong as more cards are revealed. However, one of the supplementary analyses in the SI restricts itself to betting rounds for which the speculative value is easy to calculate (Fig. S1).

[6] The SI establishes the robustness of our results within the subset of hands with fully observed state.

    2.1.    Cheating is not unheard of — there have been well-publicized scandals in online poker[7] — but we assume it away here.

    2.2.    There is technically an effect of *P1* on *W2* via *P2*. *P1* and *P2* share a statistical dependence because cards are dealt without replacement; if one player holds the Ace of Hearts, then no other player holds that Ace. This dependence is weak and we assume it away.

    2.3.    The *W1→W2* channel does not have the capacity to transmit more than one variable's worth of information (one symbol) per unit time. *P1* cannot contribute to synergy with *W1* about *W2* if its only access to *W2* is via *W1,* it requires a second channel.

3.  Looking outside of the three variables *W1*, *P1*, and *W2*, our results may also be contaminated by exogenous variables with causal influences on both *W1* and *W2*. A factor that affects both of these variables may register spurious synergy between the hero's cards and the villain's wagers. We account for those unobserved variables by generalizing the interpretation of *W2* to include such exogenous factors. Information that affects both *W1* and *W2* is in some sense public, and so we interpret *W2* as a proxy not just for information in the villain's wagers, but for all public information relevant enough to influence players' wagers. These sources include the community cards, previously observed behavior, and common knowledge about either player. This is the sense in which we identify *W2* as representing public information generally.

---

[7] See, for example, http://www.nbcnews.com/id/21381022/#.VHikx5PF9q0

4. Rather than relying strictly on logical points to support causal inferences from *W2* to *W1*, we also designed our statistical analysis to control for their mutually causal effects at each of the three scales of contamination: within betting rounds, within hands, and across hands. We ran a total of four analyses on different subsets of the data to establish the robustness of our results to conditions of greater statistical independence (within more biased samples) on the one hand, and conditions of greater statistical dependence (within less biased samples) on the other. The supporting analyses are detailed and compared in the SI.

## Acknowledgements

The authors would like to thank Asaf Beasley and Dirk Helbing and his lab for their constructive insights.

## Data accessibility

Our analysis was based on a publicly available dataset. The details of access are available in the Data subsection of the Methods.

**Supplementary Information for**

**Information encryption in the expert management of strategic uncertainty**

Seth Frey[a,b,1], Paul L. Williams[b], Dominic K. Albino[c,2]

[a] Disney Research Zürich, Stampfenbachstrasse 48, 8006, Zürich, Switzerland
[b] Cognitive Science Program, Indiana University, 1900 E 10th St., Bloomington, IN, 47406, USA
[c] Department of Economics, University of Connecticut, 365 Fairfield Way, Storrs, CT 06269, USA

———————————

[1] Correspondent. Email: seth@disneyresearch.com; Phone: +41 44 632 7360

[2] Former professional No-Limit Hold'em cash game player.

### ***Specific information decomposition***

Our central results are based on information about *W1*, as averaged over states signifying wagers of zero, small wagers, and large wagers. However, insofar as poker strategy depends on context, different variables and processing operations will be more and less important for informing different kinds of wagers. Furthermore, averages over more common events (like a small wager) can drown out interesting behavior during less common ones (like wagering zero). Fortunately, as shown in Fig. 3, it is possible to decompose the total information more finely, revealing the roles of synergy, redundancy, and unique information *within* each level of wager (Beer & Williams, 2014; Williams, 2013).

There is a *specific* analogue for every facet of the information decomposition. In three variables, the specific total information is

$$I(X = x; Y_1, Y_2) = \sum_{y_1} \sum_{y_2} p(y_1, y_2 | x) \log \frac{p(y_1, y_2 | x)}{p(y_1, y_2)}$$

The specific redundancy is

$$Rdn(X = x; Y_1, Y_2) = min(I(X = x; Y_1), I(X = x; Y_2))$$

The specific unique information from $Y_1$ is

$$Unq(X = x; Y_1) = I(X = x; Y_1) - Rdn(X = x; Y_1, Y_2)$$

and the specific synergy is

$$Syn(X = x; Y_1, Y_2) = I(X = x; Y_1, Y_2) - Unq(X = x; Y_1) - Unq(X = x; Y_2) - Rdn(X = x; Y_1, Y_2)$$

These terms have analogous interpretations to their coarser counterparts. For example, the specific synergy *Syn(W1 = "Large wager"; P1, W2)* is the information about the hero's large wagers that is contained in the interaction between the hero's cards and the villain's wagers.

***Robustness of results after controlling for temporal dependencies and with speculative hand values.***

The validity of our results requires that no causal influences between *W1*, *W2,* and *P1* flow *from W1*. But implicit in our design choice to bin all betting rounds together is the assumption that these rounds are statistically independent of each other. Consequently, our main analysis is vulnerable to the possibility that *W1* at one betting round has a systematic effect on *W2* on a subsequent round. We report here a supplementary analysis that reproduced our results on a subset of the data with fewer temporal dependencies. This analysis also had more accurate hand valuations. We controlled for within-hand statistical dependencies by repeating our analyses over only the first round of betting from each hand (the "preflop action"). This filtering step improves the statistical independence of observations in our sample because it reduces the density of our data to one datum per hand, and, in particular, because the first round of betting in a hand cannot have been influenced by behavior in any other part of that hand. By removing three out of four intermediary betting rounds, this step reduces the sensitivity of our analysis to any "medium range" statistical dependencies (direct effects of betting round *t* on rounds *t+1*, *t+2*, and *t+3*) that may persist across consecutive hands played by the same people.

We recalculated the partial information decomposition and the specific partial information decomposition over only the first round of betting, after players have received their hole cards but before any community cards have been revealed. As can be observed in Fig. S1, these supplementary results are the same as those in our core analysis: fish are more likely to cohere with unique information from *W2*, while sharks show greater coherence with the synergy between *P1* and *W2*.

One additional, striking feature of Fig. S1 is that it registers non-zero values of unique information between *W1* and *P1*. This may mean that the important aggregate influence of *W2* on *W1* does not manifest itself in the earliest betting round, only later.

This supplementary robustness check, focused as it is on the first betting round, also functions to check another assumption underlying our main analysis. Our main results may be vulnerable to one of the details of how we discretized *P1*. In rating the strengths of players' hands, we did not account for their "speculative value." Two hands that are equally bad at a given moment may still differ in their likelihood of eventually becoming strong, as more community cards are revealed. This likelihood is difficult to calculate in general, but it is relatively simple to calculate in the first betting round, when each player has only two private cards and there are no community cards. Our ranking of two-card hands in the first betting round is based on a widely accepted ordinal ranking of the 169 distinguishable levels of hole cards strength, one that accounts exhaustively for their potential to become strong.[1] The results of Fig. S1 support the proposition that our central results persist under this more controlled valuation of hand strengths.

### ***Robustness under conditions of complete information***

As elaborated in the methods, hands in our main dataset included information about players' hole cards only when that hand ended in cards being revealed during showdown. Since most hands conclude with neither player revealing their private hole cards, most values of *P1* are coded as *Not observed*. Players only reveal their cards if more than one player persists through all four betting rounds to showdown. And even then, it may be that the only player to reveal any cards is the winner — at many houses the losing players may "muck," or discard their cards without revealing them. Since our analysis defines the hero to be the second player, mucking

---

further constrains the conditions under which the state of *P1* equals something besides *Not observed* (which, in our dataset, describes *P1* in about 85% of hands). The logical structure of our argument is such that missing data should not undermine it. However, as a check for robustness, we reanalyzed the 15% of rounds for which complete information was available.

Why should our analysis be robust to about 85% missing data? Our core claim is that poker sharks are more likely to process public information in a way that obscures its effects on their behavioral patterns. This claim depends on our observation that levels of synergy between *P1* and *W2* are greater among sharks than fish. If our estimate of synergy is anti-conservative — if missed observations of true hand strength make us likely to detect spurious synergies — then the prevalence of the *Not observed* state would threaten our result. Conversely, if, as we believe, missing observations of *P1* make us likely to uniformly underestimate synergy, then our results should be robust to high frequencies of folded hands

This supplementary analysis reproduces our main observations, that fish's wagers cohere more with unique information in *W2*, and that shark wagers depend more on the synergy between *W2* and *P1*. According to Fig. S2, our core results hold in this subset of the data, though only in the case of large wagers (Third row of the specific information decomposition). While a decomposition that averages over all states of *W1* finds no significant differences between sharks and fish (Top row of S2), the specific decomposition shows that this lack of an average effect is due to high levels of noise for the *No wager* state of *W1* (First row of the specific information decomposition). This check is necessarily more narrow. Standing alone, it holds only for the subset of cases in which the hero was in second position and won a hand by revealing cards at showdown. But, as complemented by the corroborative results of the other robustness checks, this supplementary analysis supports our central claims that sharks cohere more with the synergy between information sources and less with *W2* uniquely.

### *Robustness to player order and sequential dependence*

This robustness check modifies the main analysis by sampling from the behavior of either the first or the second player in each betting round. We present ourselves as informing the actions of poker experts generally, not just poker experts who play in the second position, but none of the above analyses control for the possibility, however remote, that synergy is not due to player expertise, but to player position. Indeed, a player's position at the table may have important effects on their strategy. The recent result by Bowling et al. prove that, in two-person Limit Hold'em, the player in second position has an advantage of almost one-hundredth of the (big) blind (2015). Additionally, if expert players somehow manage to play more hands in second position than do novices — perhaps by switching tables less often — then skill and position are confounded and our results may be due to the latter.

At first glance, including the first player in the analysis violates the assumptions that permit the causal influence of *P1* and *W2* on *W1*. Treating the first player as the focal player in a causal analysis seems to require that that player is responding to actions that have not happened yet. However, reexamining the assumptions section of the Methods will show that averaging over play order does not actually violate the assumptions we used to justify a causal interpretation of the results. Even when *W2* occurs after *W1*, any synergy or redundancy *about* one of these three variables *from* the other two must be about *W1* from *P1* and *W2*, since *W1* is the only of the variables with causal inputs from both of the others. How could it be justified to treat *W2* as an input to *W1* before it occurs? If sharks are substantially better at predicting the villain's wagers, it may be that their behavior is satisfactorily modeled with an assumption that the villain's behavior has already occurred. Alternatively, the statistical dependence between betting rounds (which we are vulnerable to here, but that we control for in the other robustness analyses) may also provide a route by which the hero gains information about *W2* before it has been observed.

The results of this last robustness check show a surprising consonance with those of the main analysis (compare Figs. 2 and 3 with Fig. S3). The only major difference is in the middle row of the specific partial information decomposition, which shows that the hero's small wagers (less than or equal to the blind) are predicted by different patterns of information depending on their position at the table. Position does not seem to have a strong effect on whether the hero will post no wager or a large wager.

**Table S1. Data summarized by betting level.**

| betting level | # hands | # hands 2 players | binning cutoff: max small wager | binning cutoff: max weak hand | binning cutoff: max mean earnings of a fish | % sharks | $p_{fish}(W1) =$ {0, Small, Large} $p_{shark}(W1) =$ {0, Small, Large} | $p_{fish}(W2) =$ {0, Small, Large} $p_{shark}(W2) =$ {0, Small, Large} | $p_{fish}(P1) =$ {NA,Weak,Strong} $p_{shark}(P1) =$ {NA,Weak,Strong} |
|---|---|---|---|---|---|---|---|---|---|
| $0.25 | 4'045'738 | 62'149 | $ 0.25 | pair 4s | −0.4 | 14.9% | {0.080 0.483 0.437} | {0.108 0.514 0.378} | {0.799 0.111 0.090} |
|  |  |  |  |  |  |  | {0.092 0.557 0.351} | {0.097 0.557 0.346} | {0.803 0.102 0.094} |
| $0.50 | 1'216'576 | 611'386 | $ 0.50 | pair 3s | −0.64 | 13.3% | {0.123 0.506 0.371} | {0.094 0.465 0.441} | {0.806 0.090 0.104} |
|  |  |  |  |  |  |  | {0.128 0.536 0.336} | {0.072 0.496 0.432} | {0.774 0.102 0.124} |
| $1.00 | 1'323'528 | 454'246 | $ 1.00 | pair 4s | −1.24 | 12.9% | {0.144 0.510 0.346} | {0.098 0.477 0.425} | {0.839 0.072 0.089} |
|  |  |  |  |  |  |  | {0.144 0.512 0.344} | {0.077 0.501 0.422} | {0.807 0.084 0.110} |
| $2.00 | 1'189'242 | 337'432 | $ 2.00 | pair 5s | −2.43 | 11.9% | {0.146 0.509 0.344} | {0.103 0.470 0.427} | {0.846 0.068 0.086} |
|  |  |  |  |  |  |  | {0.159 0.486 0.355} | {0.081 0.497 0.422} | {0.825 0.074 0.101} |
| $4.00 | 845'455 | 140'858 | $ 4.00 | pair 6s | −4.80 | 11.0% | {0.152 0.493 0.355} | {0.109 0.471 0.419} | {0.858 0.060 0.082} |
|  |  |  |  |  |  |  | {0.155 0.490 0.355} | {0.097 0.498 0.405} | {0.843 0.062 0.094} |
| $6.00 | 332'841 | 57'057 | $ 6.00 | pair 6s | −7.33 | 12.2% | {0.153 0.494 0.353} | {0.124 0.465 0.411} | {0.876 0.051 0.073} |
|  |  |  |  |  |  |  | {0.157 0.491 0.352} | {0.093 0.506 0.402} | {0.836 0.069 0.095} |
| $10.00 | 320'686 | 87'105 | $ 10.00 | pair 6s | −12.60 | 10.0% | {0.162 0.492 0.345} | {0.121 0.468 0.411} | {0.881 0.050 0.069} |
|  |  |  |  |  |  |  | {0.156 0.502 0.343} | {0.105 0.503 0.392} | {0.849 0.062 0.090} |

Statistics about each betting level (or blind) including details of the discretization of *W1*, *P1*, and *W2* at each betting level. Note that, although most games were 25¢ games, two-player hands are unusually rare at 25¢ stakes. Consequently, they are relatively poorly represented in our dataset. From the distributions printed in the table, one may observe that wagers of zero are more common at higher (versus lower) stakes, among second- (versus first-) position players, among sharks (versus fish) in the second position, and among fish (versus sharks) in the first position.

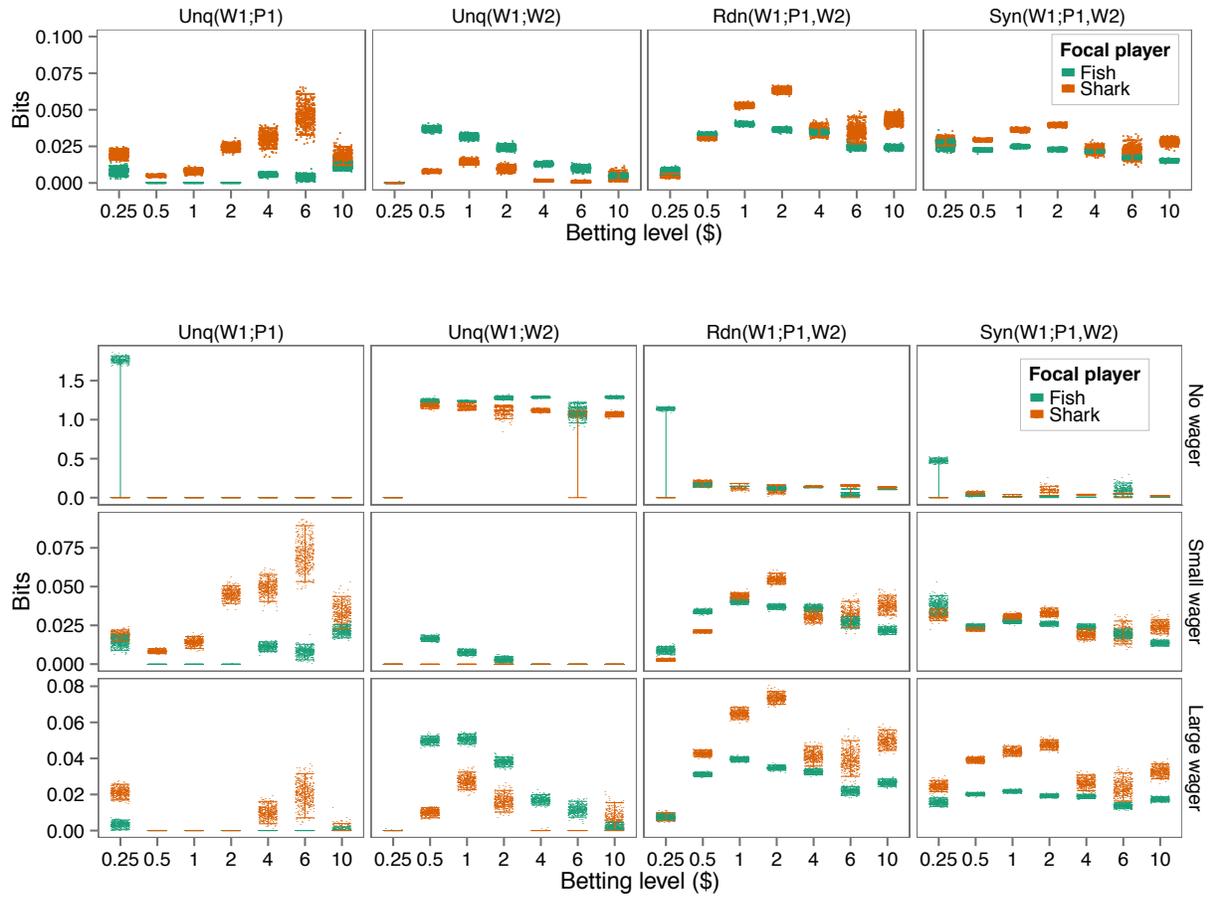

**Figure S1. Information decompositions of the total information, I(W1;P1,W2), restricted to the first betting round.** Sharks exhibit significantly more synergy and less unique information from *W2*, particularly when they made large wagers. Their wagers also exhibit non-zero contributions of unique information from *P1*.

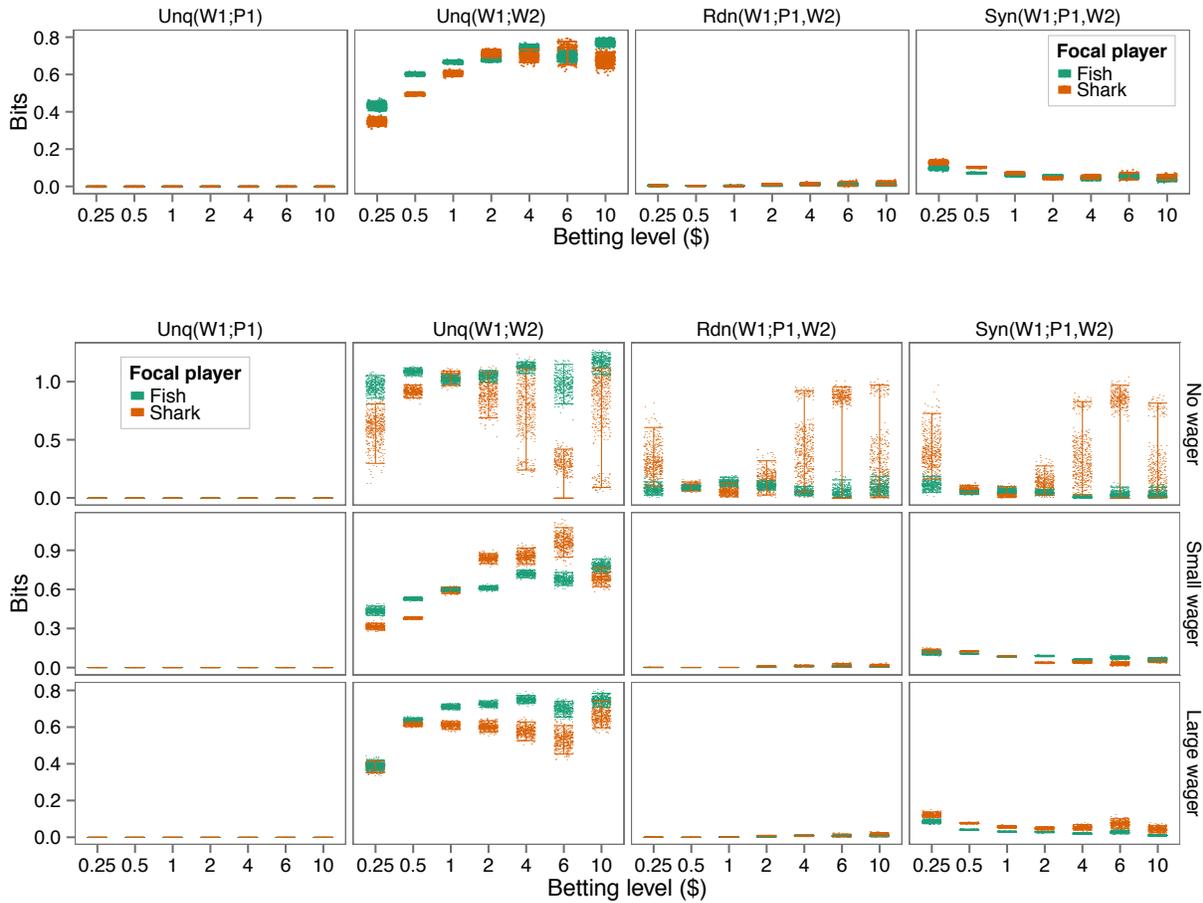

**Figure S2. Information decompositions of the total information, I(W1;P1,W2), after discarding all missing observations of *P1*.** In the case of large wagers, sharks exhibit significantly less unique information from *W2* and — though the effect is small — significantly more synergy. Averaging over the states of *W1* erases these effects, probably because of the very high variance of the results in the top row of the specific information decomposition, for *W1* = "No wager."

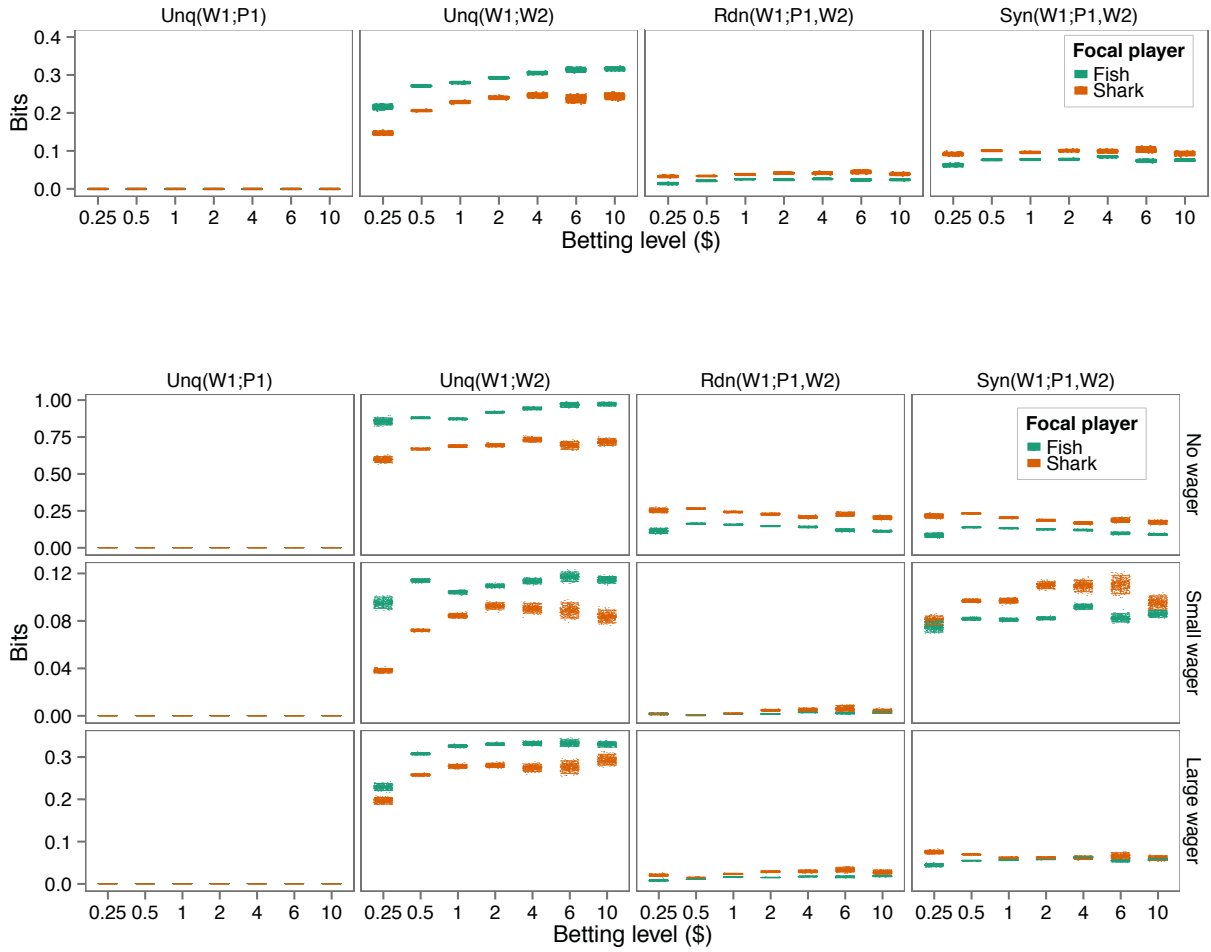

**Figure S3. Information decompositions of the total information, I(W1;P1,W2), averaging over player order.** The results of this third analysis are difficult to distinguish from those of Figs. 2 and 3, and they support the same conclusions. The only clear difference in information management caused by changes in player position seems to be in influences on small wagers, those less than or equal to the blind. As in the main analysis, sharks extract more information from *P1* and *W2* redundantly, and from their synergy, and fish extract more information from *W2* uniquely.